\documentclass[12pt]{article}

\usepackage{graphicx}
\usepackage{cite}
\usepackage{amsmath}
\usepackage{amssymb}

\begin{document}

\hoffset = -1truecm \voffset = -2truecm \baselineskip = 6 mm

\title{\bf Applications of a nonlinear evolution equation II: the EMC effect}

\author{
{\bf Xurong Chen}$^1$, {\bf Jianhong Ruan}$^2$, {\bf  Rong Wang}$^{1,3}$\\
{\bf Pengming Zhang}$^1$ and {\bf Wei Zhu}$^2$\footnote{Corresponding author, E-mail: weizhu@mail.ecnu.edu.cn}\\
\\
\normalsize $^1$Institute of Modern Physics, Chinese Academy of Sciences, Lanzhou 730000,
 P.R. China\\
\normalsize $^2$Department of Physics, East China Normal University,
Shanghai 200062, P.R. China \\
\normalsize $^3$ University of Chinese Academy of Sciences, Beijing, 100049, P.R. China
}

\date{}

\newpage

\maketitle

\vskip 3truecm

\begin{abstract}

    The EMC effect is studied by using the DGLAP equation with the ZRS corrections and
minimum number of free parameters, where the nuclear shadowing
effect is a dynamical evolution result of the equation, the
nucleon swelling and Fermi motion in the nuclear environment deform
the input parton distributions. Parton distributions of both proton and nucleus are predicted in
a unified framework. We show that the parton recombination as a
higher twist correction plays an essential role in the evolution of
parton distributions either of proton or nucleus. We find that the
nuclear anti-shadowing contributes a part of
enhancement of the ratio of the structure functions around $x\sim
0.1$, while the other part origins from the deformation of
the nuclear valence quark distributions. In particularly, the nuclear gluon distributions are dynamically
predicted, which are important information for the recherche of the high energy nuclear physics.

\end{abstract}

PACS number(s):13.60.Hb; 12.38.Bx

$keywords$: EMC effect, Nuclear shadowing and anti-shadowing, QCD
evolution equation; Nonlinear corrections

\newpage
\begin{center}
\section{Introduction }
\end{center}

    The question that how the properties of hadrons bounded in nuclear medium
differ from that of free hadrons is an important and active research
topic of experiment and theory. Example of such medium modifications
is the European Muon Collaboration (EMC) effect in the
deep-inelastic scattering (DIS), which has been discussed
extensively since early 1980s starting from the observation of a
change in the structure function of a heavy nucleus relative to that
of the deuteron \cite{1}. The nuclear effects in DIS were further
measured in the form of the ratio $R(A/B)= F_2^A/F_2^B$ of two
nuclei. From the studies of data on the ratio R(A/B) one can divide
a few regions of characteristic nuclear effects \cite{2}: depletion
of nuclear structure functions at small Bjorken variable $x \lesssim
0.05$ (shadowing region); a small enhancement of nuclear structure
functions for $0.05 \lesssim x \lesssim 0.3$ (anti-shadowing);
depletion with a minimum around $x = 0.6\sim 0.7$ (known as EMC
effect) followed by a rise at large $x$ (Fermi motion).

    A quantitative understanding of the EMC effect can help us to understand how the
properties of hadrons are modified in a nuclear medium and even can
provide valuable insights into the origin of nuclear force.  In
particular, the EMC effect has been taken as fundamental basis in
establishing the nuclear parton distributions, which often serve as
the source of information about quark-gluon plasma (QGP) and color
glass condensation (CGC) in heavy ion collisions at the BNL
Relativistic Heavy Ion Collider (RHIC) and the CERN Large Hadron
Collider (LHC). For these sakes, a proper theory of the EMC effect
should explain not only the ratios of the structure functions but
also the absolute values of the nuclear structure functions
themselves.

    The complexity of the EMC effect is due to the fact that
the parton distributions of a bound nucleon are affected through
both perturbative and nonperturbative ways. The parton recombination
is generally thought as a source of the nuclear shadowing, which can
be described in a perturbative framework.  Close, Qiu and Roberts
described the nuclear shadowing using a perturbative QCD calculation
of the gluon  recombination function at a fixed $Q^2$ scale in
\cite{3}. This work is irrelevant to any dynamical equation and is
less in the prediction power. Mueller and Qiu used the DGLAP
equation with the GLR-MQ corrections discussed qualitatively the
nuclear shadowing in \cite{4,5}. In our previous work \cite{6} we
obtained the parton distribution functions of the proton, which are
evaluated dynamically using the DGLAP equation with the ZRS
corrections \cite{7,8,9} from a low scale, where the nucleon
consists valence quarks.  The simple input distributions of the
proton can describe the nonperturbative nuclear corrections.  The
above scheme allows us to dynamically and quantitatively research
the EMC effect. For this sake, in this paper we first generalize the
ZRS corrections to the deep inelastic scattering off a nuclear
target in Sec. 2. We find that the nuclear shadowing effect is a
dynamical result of the parton recombination due to multi-nucleon
correlations in the nuclear target. Then we add the contributions of
a traditional Fermi motion and the nucleon swelling on the nuclear
input parton distributions in Secs. 3 and 4, respectively. A series
of the EMC effect ratios are presented in Sec. 5, which are
consistent with the present experimental data. We also calculate the
anti-shadowing effect and point out that the so-called
anti-shadowing effect is mainly due to the deformation of the
valence quarks in a swelling nucleon rather than a compensation to
the shadowing correction.  In particularly, the nuclear gluon
distributions are dynamically predicted in Figs. 6 and 7, which are
important information in studying high energy nuclear physics. The
discussions and summary are given in the Sec. 6.

\newpage
\begin{center}
\section{Dynamical nuclear shadowing}
\end{center}

    The corrections of parton recombination to the QCD evolution of the parton distributions in
proton have been studied with the ZSR equation. In nuclear
target, the longitudinal localization size of a parton with small
$x$ could exceed a nucleon size. In this case, the partons in
different nucleons could interact and they contribute a factor
$A_{eff}$ in the following modified DGLAP equation with the ZRS corrections \cite{7,8,9}. We
define that $f_{v_j}^A(x,Q^2)$ (j=u,d) are valence quark
distributions, $f^A_{q_i}(x,Q^2)$ (i=u,d,s) are sea quark
distributions, $f^A_{\overline{q}_i}(x,Q^2)$ (i=u,d,s) are anti-sea
quark distributions and $f_g^A(x,Q^2)$ is gluon distribution,
$\Sigma^A(x,Q^2)\equiv\sum_j f_{v_j}^A(x,Q^2)+\sum_i
f_{q_i}^A(x,Q^2)+\sum_if_{\overline{q}_i}^A(x,Q^2)$. Thus, the
DGLAP equation with the ZRS corrections in nuclear target reads

$$Q^2\frac{dxf_{v_j}^A(x,Q^2)}{dQ^2}$$
$$=\frac{\alpha_s(Q^2)}{2\pi}\int_x^1\frac{dy}{y}P_{qq}(z)xf_{v_j}^A(y,
Q^2)$$
$$-\frac{\alpha_s(Q^2)}{2\pi}xf_{v_j}^A(x,Q^2)\int_0^1 dzP_{qq}(z)$$
$$-A_{eff}\frac{\alpha_s^2(Q^2)}{4\pi R^2 Q^2}\int_x^{1/2}\frac{dy}{y}xP_{qg\rightarrow q}(x,y)yf_g^A(y,Q^2)yf_{v_j}^A(y,Q^2)$$
$$+A_{eff}\frac{\alpha_s^2(Q^2)}{4\pi R^2 Q^2}\int_{x/2}^x\frac{dy}{y}xP_{qg\rightarrow q}(x,y)yf_g^A(y,Q^2)yf_{v_j}^A(y,Q^2)$$
$$-A_{eff}\frac{\alpha_s^2(Q^2)}{4\pi R^2 Q^2}\int_x^{1/2}\frac{dy}{y}xP_{qq\rightarrow q}(x,y)y[\Sigma^A(y,Q^2)-f_{v_j}^A(y,Q^2)]
yf_{v_j}^A(y,Q^2)$$
$$+A_{eff}\frac{\alpha_s^2(Q^2)}{4\pi R^2 Q^2}\int_{x/2}^x\frac{dy}{y}xP_{qq\rightarrow q}(x,y)y[\Sigma^A(y,Q^2)-f_{v_j}^A(y,Q^2)]
yf_{v_j}^A(y,Q^2),(if~x\le 1/2), $$

$$Q^2\frac{dxf_{v_j}^A(x,Q^2)}{dQ^2}$$
$$=\frac{\alpha_s(Q^2)}{2\pi}\int_x^1\frac{dy}{y}P_{qq}(z)xf_{v_j}^A(y,
Q^2)$$
$$-\frac{\alpha_s(Q^2)}{2\pi}xf_{v_j}^A(x,Q^2)\int_0^1 dzP_{qq}(z)$$
$$+A_{eff}\frac{\alpha_s^2(Q^2)}{4\pi R^2 Q^2}\int_{x/2}^{1/2}\frac{dy}{y}xP_{qg\rightarrow q}(x,y)yf_g^A(y,Q^2)yf_{v_j}^A(y,Q^2)$$
$$+A_{eff}\frac{\alpha_s^2(Q^2)}{4\pi R^2 Q^2}\int_{x/2}^{1/2}\frac{dy}{y}xP_{qq\rightarrow q}(x,y)y[\Sigma^A(y,Q^2)-f_{v_j}^A(y,Q^2)]
yf_{v_j}^A(y,Q^2),(if~1/2\le x\le 1), \eqno(1-a)$$ for valence
quarks, where $z=x/y$, the factor $1/(4\pi R^2)$ is from normalizing
two-parton distribution, R is the correlation length of two initial
partons,

$$Q^2\frac{dxf_{\overline{q}_i}^A(x,Q^2)}{dQ^2}$$
$$=\frac{\alpha_s(Q^2)}{2\pi}\int_x^1\frac{dy}{y}P_{qq}(z)xf_{\overline{q}_i}^A(y,
Q^2)$$
$$-\frac{\alpha_s(Q^2)}{2\pi}xf_{\overline{q}_i}^A(x,Q^2)\int_0^1dzP_{qq}(z)$$
$$+\frac{\alpha_s(Q^2)}{2\pi}\int_x^1\frac{dy}{y}P_{qg}(z)xf_g^A(y,
Q^2)$$
$$-A_{eff}\frac{\alpha_s^2(Q^2)}{4\pi R^2 Q^2}\int_x^{1/2}\frac{dy}{y}x
P_{gg\rightarrow \overline{q}}(x,y)[ yf_g^A(y,Q^2)]^2$$
$$+A_{eff}\frac{\alpha_s^2(Q^2)}{4\pi R^2 Q^2}\int_{x/2}^x\frac{dy}{y}x
P_{gg\rightarrow \overline{q}}(x,y)[ yf_g^A(y,Q^2)]^2$$
$$-A_{eff}\frac{\alpha_s^2(Q^2)}{4\pi R^2 Q^2}\int_x^{1/2}\frac{dy}{y}x
P_{q\overline{q}\rightarrow \overline{q}}(x,y)yf_{q_i}^A(y,Q^2)yf_{\overline{q}_i}^A(y,Q^2)$$
$$+A_{eff}\frac{\alpha_s^2(Q^2)}{4\pi R^2 Q^2}\int_{x/2}^{x}\frac{dy}{y}x
P_{q\overline{q}\rightarrow \overline{q}}(x,y)yf_{q_i}^A(y,Q^2)yf_{\overline{q}_i}^A(y,Q^2)$$
$$-A_{eff}\frac{\alpha_s^2(Q^2)}{4\pi R^2 Q^2}\int_x^{1/2}\frac{dy}{y}x
P_{\overline{q} \overline{q}\rightarrow \overline{q}}(x,y)y[\Sigma^A(y,Q^2)-f_{q_i}^A(y,Q^2)]yf_{\overline{q}_i}^A(y,Q^2)$$
$$+A_{eff}\frac{\alpha_s^2(Q^2)}{4\pi R^2 Q^2}\int_{x/2}^{x}\frac{dy}{y}x
P_{\overline{q} \overline{q}\rightarrow \overline{q}}(x,y)y[\Sigma^A(y,Q^2)-f_{q_i}^A(y,Q^2)]yf_{\overline{q}_i}^A(y,Q^2)$$
$$-A_{eff}\frac{\alpha_s^2(Q^2)}{4\pi R^2 Q^2}\int_x^{1/2}\frac{dy}{y}x
P_{\overline{q}g\rightarrow \overline{q}}(x,y)yf_g^A(y,Q^2)yf_{\overline{q}_i}^A(y,Q^2)$$
$$+A_{eff}\frac{\alpha_s^2(Q^2)}{4\pi R^2 Q^2}\int_{x/2}^x\frac{dy}{y}x
P_{\overline{q}g\rightarrow \overline{q}}(x,y) yf_g^A(y,Q^2)yf_{\overline{q}_i}^A(y,Q^2),(if~x\le 1/2),$$

$$Q^2\frac{dxf_{\overline{q}_i}^A(x,Q^2)}{dQ^2}$$
$$=\frac{\alpha_s(Q^2)}{2\pi}\int_x^1\frac{dy}{y}P_{qq}(z)xf_{\overline{q}_i}^A(y,
Q^2)$$
$$-\frac{\alpha_s(Q^2)}{2\pi}xf_{\overline{q}_i}^A(x,Q^2)\int_0^1dzP_{qq}(z)$$
$$+\frac{\alpha_s(Q^2)}{2\pi}\int_x^1\frac{dy}{y}P_{qg}(z)xf_g^A(y,
Q^2)$$
$$+A_{eff}\frac{\alpha_s^2(Q^2)}{4\pi R^2 Q^2}\int_{x/2}^{1/2}\frac{dy}{y}x
P_{gg\rightarrow \overline{q}}(x,y)[ yf_g^A(y,Q^2)]^2$$
$$+A_{eff}\frac{\alpha_s^2(Q^2)}{4\pi R^2 Q^2}\int_{x/2}^{1/2}\frac{dy}{y}x
P_{q\overline{q}\rightarrow \overline{q}}(x,y)yf_{q_i}^A(y,Q^2)yf_{\overline{q}_i}^A(y,Q^2)$$
$$+A_{eff}\frac{\alpha_s^2(Q^2)}{4\pi R^2 Q^2}\int_{x/2}^{1/2}\frac{dy}{y}x
P_{\overline{q} \overline{q}\rightarrow \overline{q}}(x,y)y[\Sigma^A(y,Q^2)-f_{q_i}^A(y,Q^2)]yf_{\overline{q}_i}^A(y,Q^2)$$
$$+A_{eff}\frac{\alpha_s^2(Q^2)}{4\pi R^2 Q^2}\int_{x/2}^{1/2}\frac{dy}{y}x
P_{\overline{q}g\rightarrow \overline{q}}(x,y)yf_g^A(y,Q^2)yf_{\overline{q}_i}^A(y,Q^2),(if~1/2\le x\le 1),\eqno(1-b)$$
for sea quark distributions and

$$Q^2\frac{dxf_g^A(x,Q^2)}{dQ^2}$$
$$=\frac{\alpha_s(Q^2)}{2\pi}\int_x^1\frac{dy}{y}P_{gq}(z)x\Sigma^A(y,
Q^2)$$
$$+\frac{\alpha_s(Q^2)}{2\pi}\int_x^1\frac{dy}{y}P_{gg}(z)xf_g^A(y,
Q^2)$$
$$-f\frac{\alpha_s(Q^2)}{2\pi}xf_g^A(x,Q^2)\int_0^1dzP_{qg}(z)$$
$$-\frac{1}{2}\frac{\alpha_s(Q^2)}{2\pi}xf_g^A(x,Q^2)\int_0^1 dzP_{gg}(z)$$
$$-A_{eff}\frac{\alpha_s^2(Q^2)}{4\pi R^2 Q^2}\int_x^{1/2}\frac{dy}{y}x
P_{gg\rightarrow g}(x,y)[ yf_g^A(y,Q^2)]^2$$
$$+A_{eff}\frac{\alpha_s^2(Q^2)}{4\pi R^2 Q^2}\int_{x/2}^x\frac{dy}{y}x
P_{gg\rightarrow g}(x,y)[ yf_g^A(y,Q^2)]^2$$
$$-A_{eff}\frac{\alpha_s^2(Q^2)}{4\pi R^2 Q^2}\int_x^{1/2}\frac{dy}{y}x
P_{q\overline{q}\rightarrow g}(x,y)\sum_{i=1}^{f}[ yf_{\overline{q}_i}^A(y,Q^2)]^2$$
$$+A_{eff}\frac{\alpha_s^2(Q^2)}{4\pi R^2 Q^2}\int_{x/2}^x\frac{dy}{y}x
P_{q\overline{q}\rightarrow g}(x,y)\sum_{i=1}^{f}[ yf_{\overline{q}_i}^A(y,Q^2)]^2$$
$$-A_{eff}\frac{\alpha_s^2(Q^2)}{4\pi R^2 Q^2}\int_x^{1/2}\frac{dy}{y}x
P_{qg\rightarrow g}(x,y)y\Sigma^A(y,Q^2)yf_g^A(y,Q^2)$$
$$+A_{eff}\frac{\alpha_s^2(Q^2)}{4\pi R^2 Q^2}\int_{x/2}^x\frac{dy}{y}x
P_{qg\rightarrow g}(x,y)y\Sigma^A(y,Q^2)yf_g^A(y,Q^2),(if~x\le 1/2),$$

$$Q^2\frac{dxf_g^A(x,Q^2)}{dQ^2}$$
$$=\frac{\alpha_s(Q^2)}{2\pi}\int_x^1\frac{dy}{y}P_{gq}(z)x\Sigma^A(y,
Q^2)$$
$$+\frac{\alpha_s(Q^2)}{2\pi}\int_x^1\frac{dy}{y}P_{gg}(z)xf_g^A(y,
Q^2)$$
$$-f\frac{\alpha_s(Q^2)}{2\pi}xf_g^A(x,Q^2)\int_0^1dzP_{qg}(z)$$
$$-\frac{1}{2}\frac{\alpha_s(Q^2)}{2\pi}xf_g^A(x,Q^2)\int_0^1 dzP_{gg}(z)$$
$$+A_{eff}\frac{\alpha_s^2(Q^2)}{4\pi R^2 Q^2}\int_{x/2}^{1/2}\frac{dy}{y}x
P_{gg\rightarrow g}(x,y)[ yf_g^A(y,Q^2)]^2$$
$$+A_{eff}\frac{\alpha_s^2(Q^2)}{4\pi R^2 Q^2}\int_{x/2}^{1/2}\frac{dy}{y}x
P_{q\overline{q}\rightarrow g}(x,y)\sum_{i=1}^{f}[ yf_{\overline{q}_i}^A(y,Q^2)]^2$$
$$+A_{eff}\frac{\alpha_s^2(Q^2)}{4\pi R^2 Q^2}\int_{x/2}^{1/2}\frac{dy}{y}x
P_{qg\rightarrow g}(x,y)y\Sigma^A(y,Q^2)yf_g^A(y,Q^2),
(if~1/2\le x\le 1),\eqno(1-c)$$
for gluon distribution.  The contributions of the negative nonlinear
(shadowing) terms vanish in $0.5<x<1$. The coefficient
$A_{eff}=1+\beta(A^{1/3}-1)$; $(A^{1/3}-1)$ is
the number of shadowed nucleons. The fitting parameter
$\beta=0.21$, instead of $1$, is due to the anisotropic
two-dimensions distribution of nucleons in a Lorentz boost nucleus
and the empty space among bound nucleons. Note that Eq. (1) dynamically
produces the nuclear shadowing in quark- and
gluon-distributions.

\newpage
\begin{center}
\section{Fermi motion corrections }
\end{center}

     Generally, the EMC effect at large $x$ is identified to the effect
of the Fermi motion inside nucleus. This effect was
studied by Bodek and Ritchie \cite{10} and Frankfult and Strikman \cite{11}. The
Fermi motion smears the input quark distributions in a
bound nucleon

$$F_2^A(x,\mu^2)=\int_{y\ge x}dyf_A^N(y)F_2^N(x/y,\mu^2), \eqno(2)$$where

$$f_A^N(y)=\frac{3m_N}{4k_F^3}[k_F^2-m_N^2(y-\eta_A)^2], \eqno(3)$$
for $\eta_A-k_F/m_N<y<\eta_A+k_F/m_N$; otherwise $f_A^N(y)=0.$ Here
$\eta_A=1-B_A/m_N$ according to Fermi gas model. The average effective mass
of the bound nucleons is $m^*_N=m_N-B_A$. The Fermi momentum $k_F^A$
and the binding energy $B_A$ take the values from
nuclear physics (Table 1).

Table 1.  The values of binding energy $B_A$  taking from \cite{12}, the
Fermi momentum taking from \cite{10} and the swelling coefficient
$\delta_A$ in Eq. (9).
\begin{center}
\begin{tabular}{c|c|c|c}
    \hline
    Nuclei                     &  Binding Energy $B_A$(MeV)    &  $k_F$($\mathrm{fm^{-1}}$) &  \quad $\delta_A$ \\\hline
    $\mathrm{{}^2\!H} $        &  1.10                   &       0.6         &  \quad 0.01          \\\hline
    $\mathrm{{}^4\!He}$        &  7.07                   &       1.03        &  \quad 0.055         \\\hline
    $\mathrm{{}^7\!Li}$        &  5.61                   &       0.90        &  \quad 0.030         \\\hline
    $\mathrm{{}^9\!Be}$        &  6.46                   &       1.06        &  \quad 0.055         \\\hline
    $\mathrm{{}^{12}\!C}$      &  7.68                   &       1.12        &  \quad 0.055         \\\hline
    $\mathrm{{}^{27}\!Al}$     &  8.33                   &       1.2         &  \quad 0.064         \\\hline
    $\mathrm{{}^{40}\!Ca}$     &  8.55                   &       1.27        &  \quad 0.072         \\\hline
    $\mathrm{{}^{56}\!Fe}$     &  8.79                   &       1.28        &  \quad 0.080         \\\hline
    $\mathrm{{}^{64}\!Gu}$     &  8.74                   &       1.27        &  \quad 0.083         \\\hline
    $\mathrm{{}^{84}\!Kr}$     &  8.72                   &       1.29        &  \quad 0.090         \\\hline
    $\mathrm{{}^{108}\!Ag}$    &  8.54                   &       1.31        &  \quad 0.096         \\\hline
    $\mathrm{{}^{118}\!Sn}$    &  8.52                   &       1.32        &  \quad 0.099         \\\hline
    $\mathrm{{}^{131}\!Xe}$    &  8.42                   &       1.32        &  \quad 0.101         \\\hline
    $\mathrm{{}^{197}\!Au}$    &  7.92                   &       1.35        &  \quad 0.112         \\\hline
    $\mathrm{{}^{208}\!Pb}$    &  7.87                   &       1.36        &  \quad 0.113         \\\hline
    \end{tabular}
    \end{center}

    Different from the binding model \cite{13,14,15}, we take $B_A$ as the binding energy per nucleon
($\sim$ a few MeV) according to nuclear physics (see Table 1)
rather than the separation energy (20 $\sim$ 40 MeV). It implies
that the bound nucleon is nearly on shell, since the nucleon
consists of three valence quarks at $\mu$-scale and hence without
any extra component (pion) to balance the lost momentum in
the off-mass shell effect. For detailed discussion see Ref. \cite{16,17,18,19}.

\newpage
\begin{center}
\section{Nucleon swelling  }
\end{center}

    After determining the dynamics of the EMC effect at small and large $x$ regions, we focus on the EMC effect
at the intermediate $x$ region. We will choose a model from various
present explanations about EMC effect \cite{2}.

    The EMC effect at $0.3<x<0.7$ was explained in the traditional nuclear physics
such as the binding model and pion model. However, the former was
argued that it can not explain the EMC effect \cite{20} by itself
and later was ruled out due to the limits set by Drell-Yan
measurements \cite{21}. Besides, the high-precision experimental
data on light nuclei from Jefferson Lab \cite{22} suggests that the
slope of $R(A/B)$ in the $0.3 < x < 0.7$ region depends on the
location of the struck quark within nucleus, i.e., it is a local
density effect. Therefore, the EMC effect can be described in terms
of modifications to the internal partonic structure of the nucleon
in the nuclear environment. One of such corrections is the swelling
of bound nucleon \cite{23,24}. The presence of such structures
within the bound nucleon could substantially increase the
confinement radius. A natural result is deformation of the input
parton distributions in a swelling nucleon according to the
Heisenberg uncertainty principle. Such swelling mechanism was first
discussed in a constituent quark model by Zhu and Shen in Ref.
\cite{16,17,18,19}.

     The simple three quark input distributions in Ref. \cite{6} allow us easily to determine
the nuclear deformed inputs. Particularly, the valence quark
distributions in a proton and a bound nucleon in nucleus (A) at $\mu^2$ are written
as

$$
xf_{v_u}^{p(A)}(x,\mu^2)=A_u^{p(A)}x^{B_u^{p(A)}}(1-x)^{C_u^{p(A)}},
$$

$$
xf_{v_d}^{p(A)}(x,\mu^2)=A^{p(A)}_dx^{B^{p(A)}_d}(1-x)^{C^{p(A)}_d}, \eqno(4)$$ where the distributions in the proton are fixed in Ref. \cite{6} as

$$xf_{v_{u}}^p(x,\mu^2)=24.30x^{1.98}(1-x)^{2.06},$$and

$$xf_{v_{d}}^p(x,\mu^2)=9.10x^{1.31}(1-x)^{3.80}. \eqno(5)$$

We define the dispersion of the input distributions as

$$D_u^{p(A)}=\left[\frac{1}{2}<f_{v_u}^{p(A)}(\mu^2)>_3-\left(\frac{1}{2}<f_{v_u}^{p(A)}(\mu^2)>_2
\right)^2\right]^{1/2}$$

$$D_d^{p(A)}=\left[<f_{v_d}^{p(A)}(\mu^2)>_3-<f_{v_d}^{p(A)}(\mu^2)>_2^2\right]^{1/2},\eqno(6)
$$ where $<>_n$ is the $n$-th moment
of the distributions. The dispersion $D^A$  decreases  with the increased confinement scale of the initial
valence quarks: $R_A = R_p+\delta R_A~ (\delta R_A>0)$, according to the uncertainty principle.
For $\delta R_A<<R_p$ we have the following relation

$$D^A_{u(d)}(\mu^2)/D_{u(d)}^p(\mu^2)=R_p/R_A \equiv \frac{1}{1+\delta_A}. \eqno(7)$$

    The swelling of a bound nucleon arises from the nuclear force and relates to the local
environment-the number of nucleons around it, since the nuclear
force is a short distance interaction. Thus the A-dependence of
swelling should be determined by the local properties of nuclear
matter, i.e., the local nuclear density $\rho_A(r)$ ($\int
\rho_A(r)d^3r=A$). We adopt the
following simple way to estimate the swelling coefficient for
$A>12$

$$\delta_A=[1-P_s(A)]\delta_0+P_s(A)\delta_0/4,\eqno(8)$$ where $\delta_0$
is the swelling coefficient for a nucleon in the nuclear
center. The second term is for the fact that the swelling of a
nucleon on the nuclear surface is less than that of one in the
interior: $1/2-$factor considers the average density of nucleon on
the surface is about half of that in the nuclear kernel and the
other $1/2$-factor is due to the space surround a nucleon on the
nuclear surface, a half is vacuum and half is nuclear mater.
$P_s(A)$ is the probability of a struck nucleon locating on the
nuclear surface, and it can be estimated in terms of the nuclear
density

$$\rho_A(r)=\rho_0/[1+\exp[(r-R_A)/b]],\eqno(9)$$with $\rho_0=0.17$ nucleon/fm$^3$, $b=0.54$ fm,
and $R_A=1.12A^{1/3}-0.86A^{-1/3}$. The surface thickness is given by
$D = (4\ln 5)b$. Thus,

$$P_s(A)=1-\int_0^{R_A-(D/2)}dr\rho_A(r)/A.\eqno(10)$$ According to this picture, scattering from a central or deeply bound constituent gives
a larger EMC effect than the scattering from a surface or weakly
bound constituent. That is, the swelling effect is a local nuclear
dynamical one. The free parameter $\delta_0=0.217$ is determined by  the
EMC ratio of any nucleus, for example, $F_2^{Ca}/F_2^D$.

\newpage
\begin{center}
\section{The EMC effect}
\end{center}

    The EMC effect is exhibited in the ratio of the DIS structure function per nucleon
bounded in a nucleus relative to that of deuterium $D$.  In this work the
isospin asymmetry is neglected. Therefore, we use $F_2^{2H}$ to
replace $F_2^D$, and the related parameters for $2H$ are listed in
Table 1, where the $k_f^{2H}$ is regarded as a parameter rather than
the Fermi momentum since the gas model is unavailable. The ratio
$R=F_2^{2H}/F_2^p$ is shown in Fig. 1.

\begin{figure}[htp]
\centering
\includegraphics[width=0.55\textwidth]{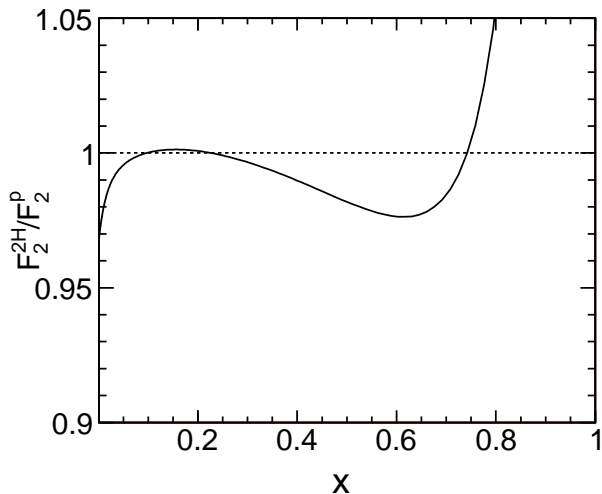}
\caption{The isoscalar EMC effect ratio $F_2^{2H}/F_2^p$ at $Q^2=5 ~$GeV$^2$.}
\label{fig1}
\end{figure}

      There are four free parameters used in the determination of the proton parton distributions
in our previous work \cite{6}. In this work we add two free parameters:
the coefficient $A_{eff}=0.210$ and $\delta_0=0.217$ to predict
the nuclear parton distributions. The other parameters in Table 1
are fixed by the knowledge of general nuclear physics.

    With the above preparation, in Fig. 2 we present $F_2^A/F_2^{2H}$
for various nuclei as a function of $x$ at $Q^2 = 5 ~$GeV$^2$. The
experimental data are taken from the publications by the European
Muon Collaboration (EMC) \cite{25,26}, the E49, E87, E139, and E140
Collaborations \cite{27,28,29,30}, the New Muon Collaboration (NMC) \cite{31,32,33,34}, BCDMS \cite{35}, HERMES
\cite{36} and Jefferson Lab \cite{22}.

\begin{figure}[htp]
\centering
\includegraphics[width=0.8\textwidth]{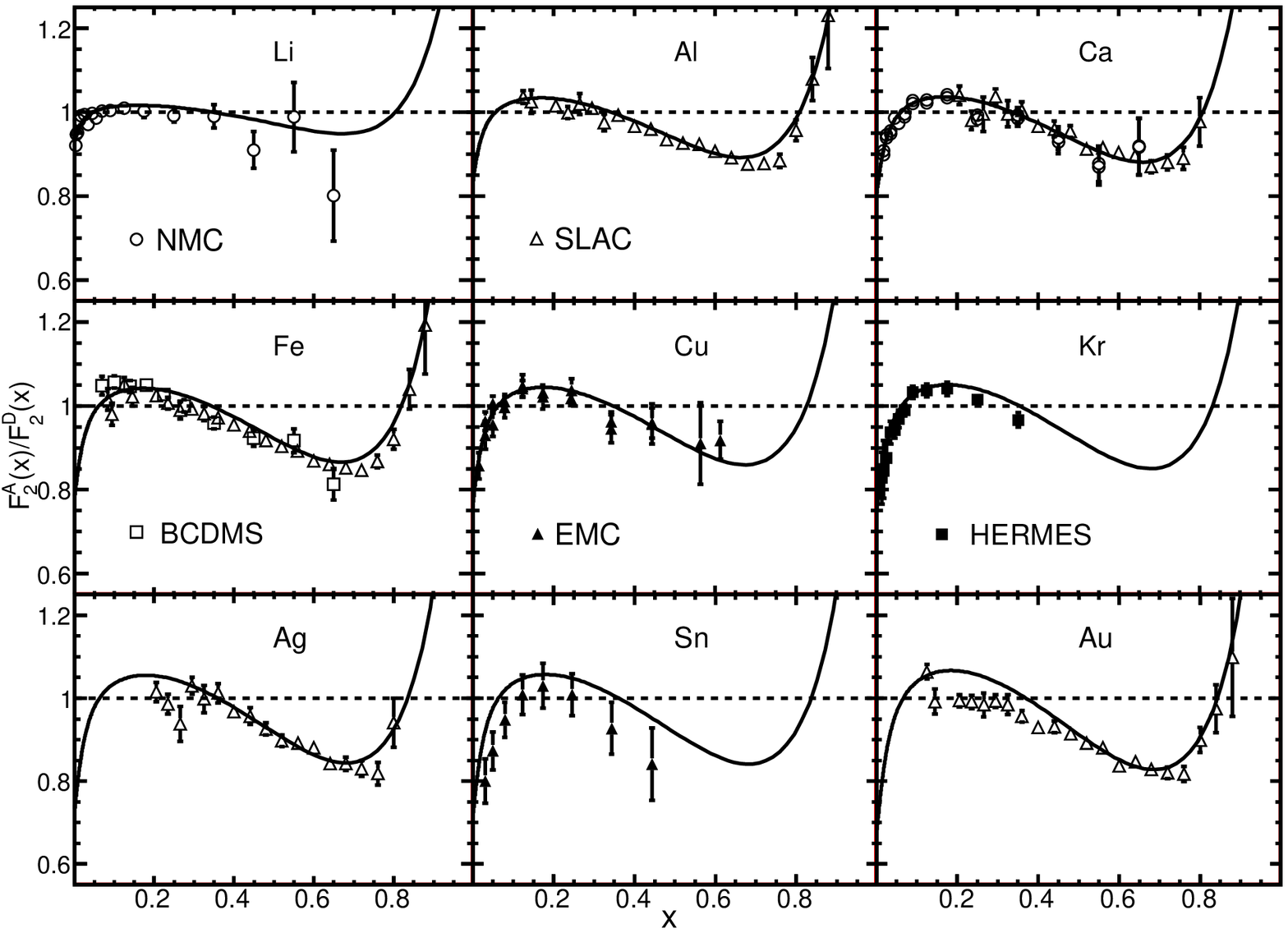}
\includegraphics[width=0.8\textwidth]{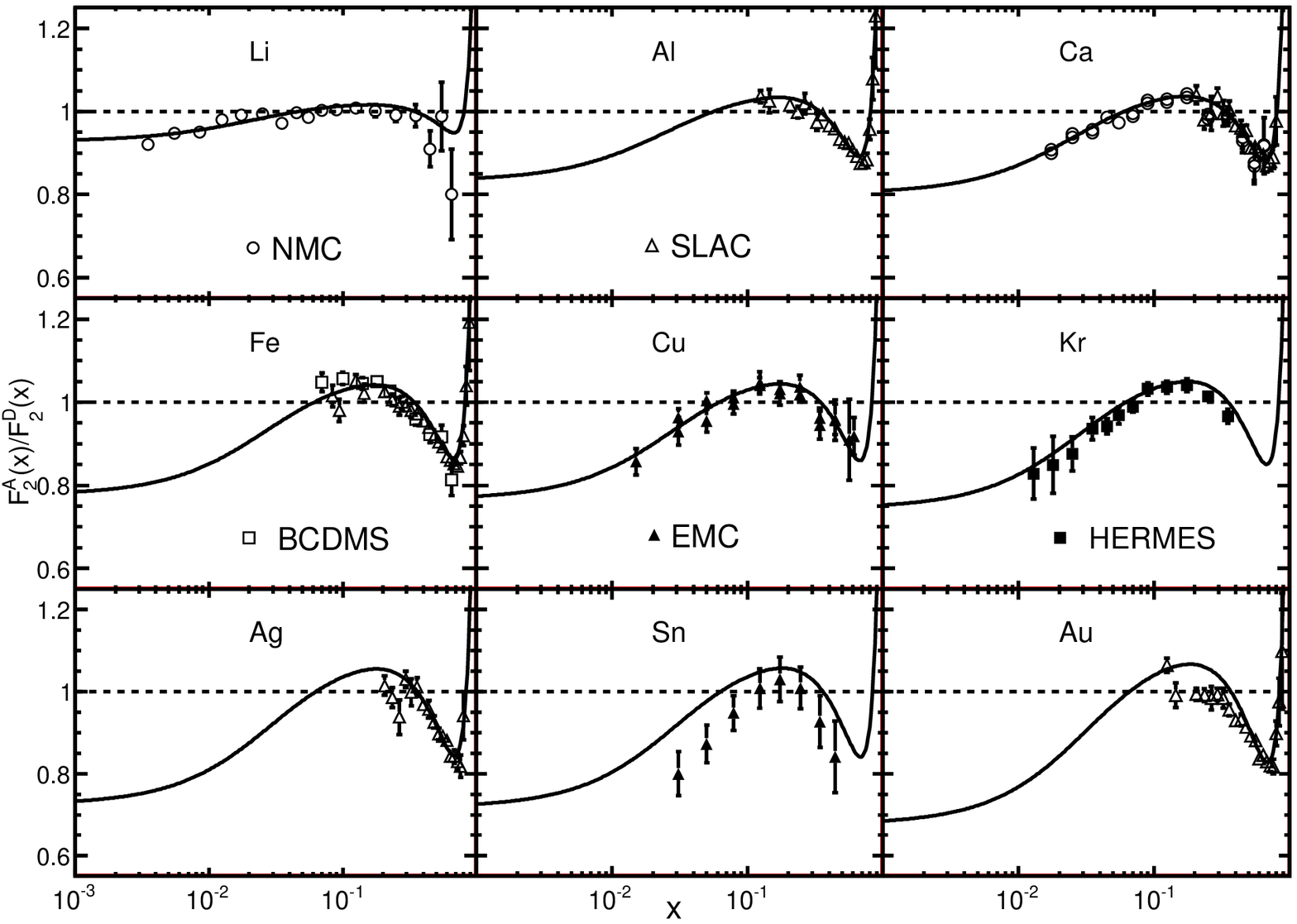}
\caption{Predicted EMC effect ratios $F_2^A/F_2^{2H}$ (solid curves)
for various nuclei at $Q^2=5~$GeV$^2$ and comparisons with
experimental data for $F_2^A/F_2^{D}$ at $Q^2>1 ~$GeV$^2$. The upper and lower figures
take different $x$ scales but for the same results.}
\label{fig2}
\end{figure}

    As it can be seen, these ratios are in good agreement with the experimental data.
The parton distributions of the proton are determined using the ZRS
equation in \cite{6}. It implies that we can not only predict the ratio
of the structure functions, but also can give the absolute values of
the nuclear structure functions themselves. These results are useful
in the research of heavy ion collisions.

    The nuclear dependence of the ratios in light nuclei $^4He$, $^9Be$ and $^{12}C$
for $x>0.2$ in Fig.3 presents that the EMC effect in this range
depends on the local nuclear environment rather than on the average
density. $^4He$ has similar average density with $^{12}C$. While
$^9Be$ is a special nucleus, which is constructed by a pair of
tightly bound alpha particles plus one additional neutron \cite{37} and
most of its nucleons are in a dense environment, similar to $^4He$.
In consequence, we set $\delta_{He}\simeq
\delta_{Be}\simeq\delta_{C}$. The clustering of nucleons in $^9Be$
leads to a special case where the average density does not reflect
the local environment of the bulk of the nucleons.

\begin{figure}[htp]
\centering
\includegraphics[width=0.65\textwidth]{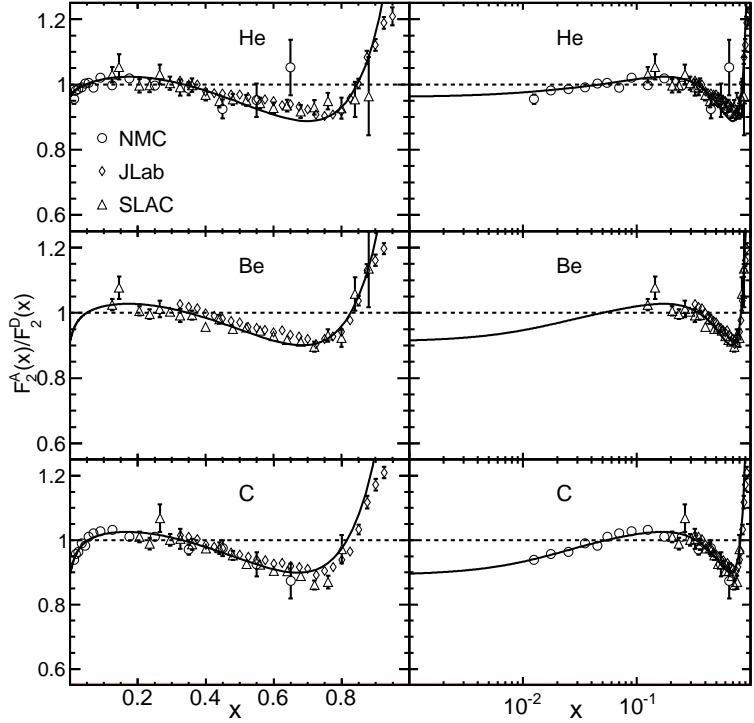}
\caption{The structure function ratios for light nuclei. The solid curve is
our theoretical predictions; (Left) and (Right) take different $x$ scales but for the same results.}
\label{fig3}
\end{figure}

    The NMC data \cite{31,32,33,34} was measured simultaneously two different nuclear targets in one experiment with systematic errors being reduced significantly. Our predictions and comparisons with the NMC data are shown
in Fig. 4. It is interesting that the ratios in Fig. 4 are finite at
$x=1$ if $k^A_F\sim k^B_F$ and it means that the ratios can be
extended to $x>1$. We call it the weak short range correlation
(wSRC) \cite{38,39,40} due to the Fermi motion.

\begin{figure}[htp]
\centering
\includegraphics[width=0.7\textwidth]{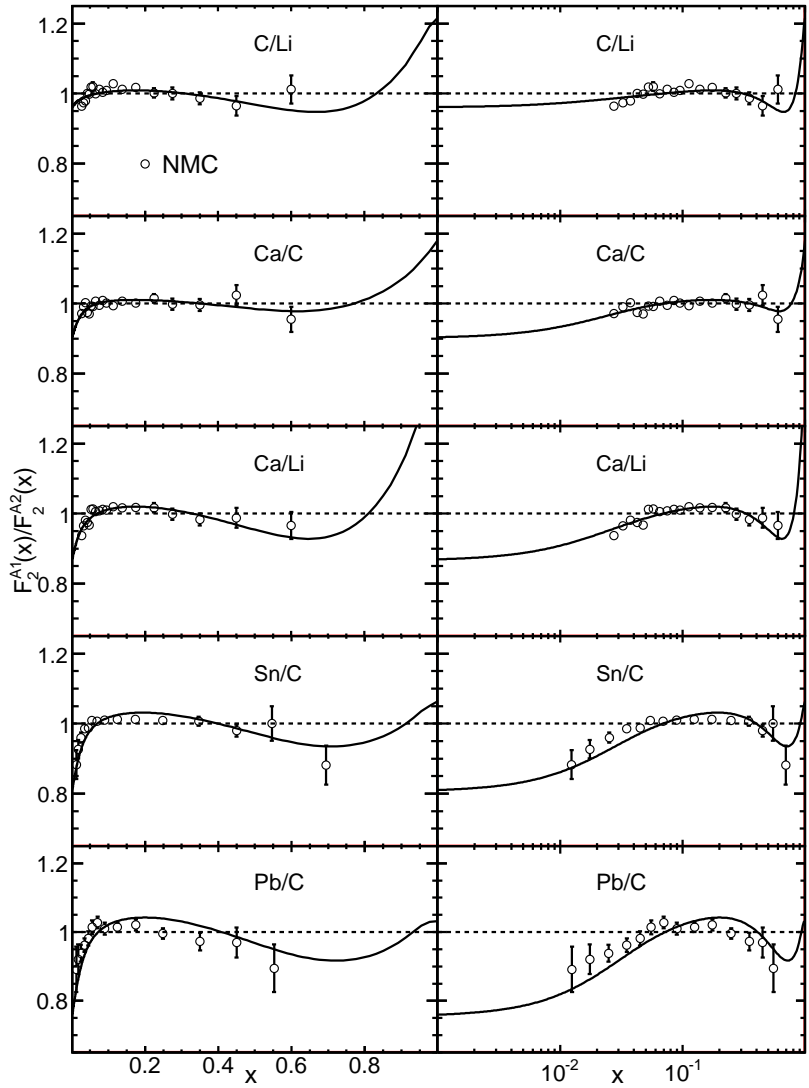}
\caption{Our predicted structure function ratios $F_2^A/F_2^{A'}$ at $Q^2=5~
$GeV$^2$ (solid curves) are compared with experimental data at $Q^2>2~ $GeV$^2$.}
\label{fig4}
\end{figure}

    Now let us discuss the structure of the EMC effect. We remove the contributions from bound nucleon
swelling and Fermi motion effects in the ratio $R=F_2^{Ca}/F_2^D$.
The remaining part in Figs. 5(a) and 5(b) is the shadowing and
anti-shadowing effect. We find that the anti-shadowing effect
contributes  only a small fraction for  the ratio in $0.05<x<0.3$
because it distributes in a broad $x$ range. Note that the momentum
conservation holds in Eq. (1) and the anti-shadowing effect is its
dynamical result.

  For the Figs. 5(a) and 5(b), when we include the swelling effect, we get Figs. 5(c) and 5(d). The swelling effect
 enhances the
distributions in the region $0.1<x<0.3$ and results in part of
"anti-shadowing". This enhancement depends on the nuclear density
(through $\delta_A$) and, furthermore, does not disappear at large
$Q^2$.

\begin{figure}[htp]
\centering
\includegraphics[width=0.6\textwidth]{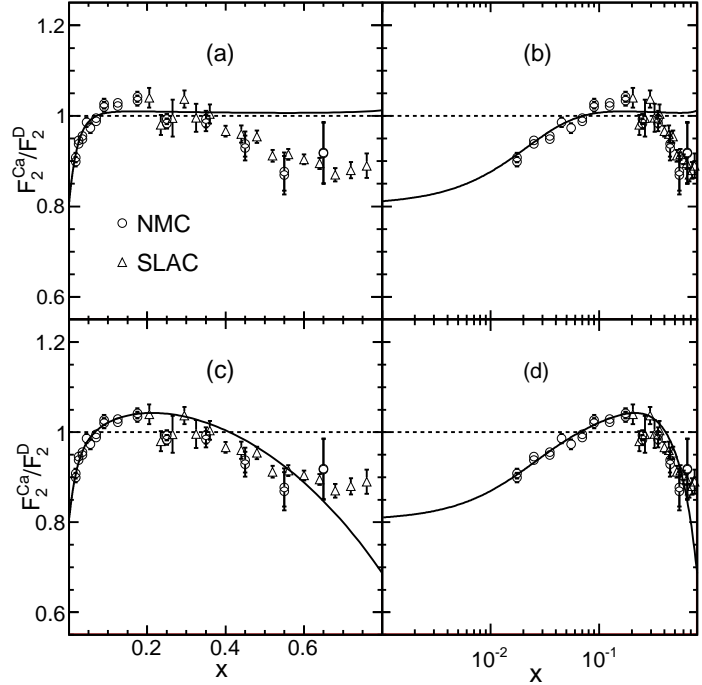}
\caption{Nuclear shadowing and antishadowing contributions to the
ratio $F_2^{Ca}/F_2^D$ at $Q^2=5~$GeV$^2$ are shown in (a) and (b);
The swelling effect is added in (c) and (d).} \label{fig5}
\end{figure}

      An example of theoretical prediction for the ratio
$xf_g^{Ca}(x,Q^2)/xf_g^p(x,Q^2)$ at $Q^2=5 ~$GeV$^2$ in the nuclear
gluon distribution is shown in Fig.6, where (a) and (b) are pure
shadowing and anti-shadowing effect, in (c) and (d) swelling effect
is added and (e) and (f) is the resulting EMC effect for gluon
including Fermi motion. We find that both nuclear shadowing  and
anti-shadowing effect in the gluon distribution are not stronger
than those in the quark distributions. In particular, we find a
complicated structure of the EMC effect in $0.05<x<0.3$, where the
anti-shadowing and swelling effects have different $x$- and
$Q^2$-dependence.

\begin{figure}[htp]
\centering
\includegraphics[width=0.65\textwidth]{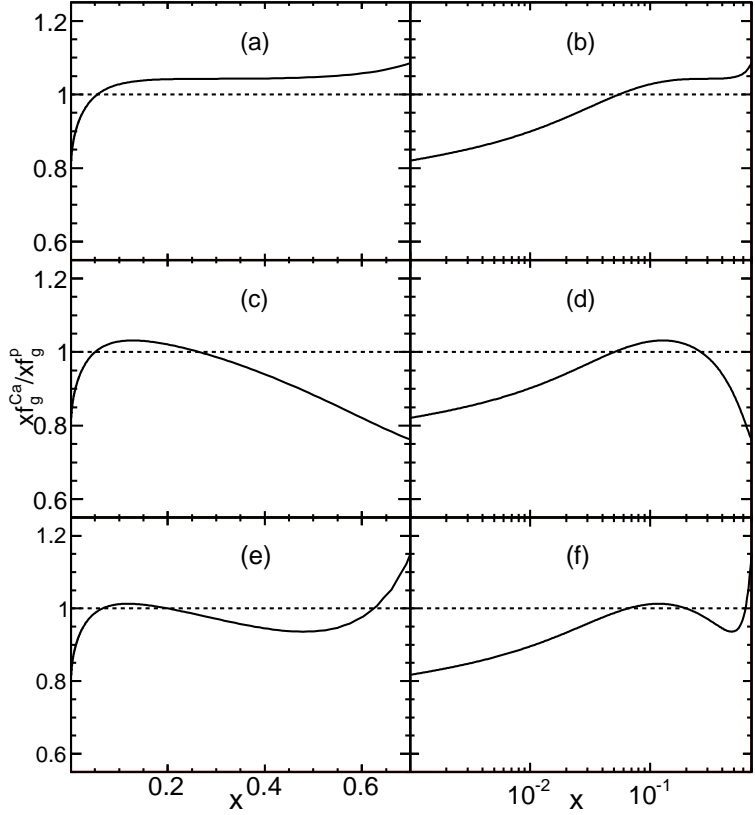}
\caption{Nuclear shadowing and antishadowing contributions to the
ratio $xf_g^{Pb}/xf_g^p$ at $Q^2=5~$GeV$^2$ are shown in (a) and
(b); The swelling effect is added in (c) and (d); A complete EMC
effect (i.e., shadowing, antishadowing, swelling and Fermi motion)
for gluon distribution ratio $xf_g^{Ca}/xf_g^p$ is shown in (e) and
(f).} \label{fig6}
\end{figure}

     Comparisons of our prediction $xf_g^{Pb}(x,Q^2)/xf_g^p(x,Q^2)$ at $Q^2=1.69 ~$GeV$^2$ for the
gluon distributions with several DGLAP-based global models
\cite{41,42,43,44} are plotted in Fig. 7. Note that in the global
analysis, the gluon distributions are mainly extracted by using the
scaling violation. However, the experimental data about the EMC
effect are restricted in a narrow $Q^2$ range at a fixed $x$. It
would be rather difficult to pin down the nuclear gluon
distributions in any global analysis. Besides, DGLAP equation works
from a larger $Q^2> 1 ~$GeV$^2$, where the $x$-dependent input
distributions contain many $ad~hoc$ free parameters. Conversely,  in
this work the gluon distributions in either proton and nucleus are
dynamically determined by Eq. (1), in which the input parameters are
fixed by the proton structure functions and only a few extra free
parameters are added for nucleus. One can expect that the
uncertainty of nuclear gluon distributions in our work will be much
reduced.

\begin{figure}[htp]
\centering
\includegraphics[width=0.55\textwidth]{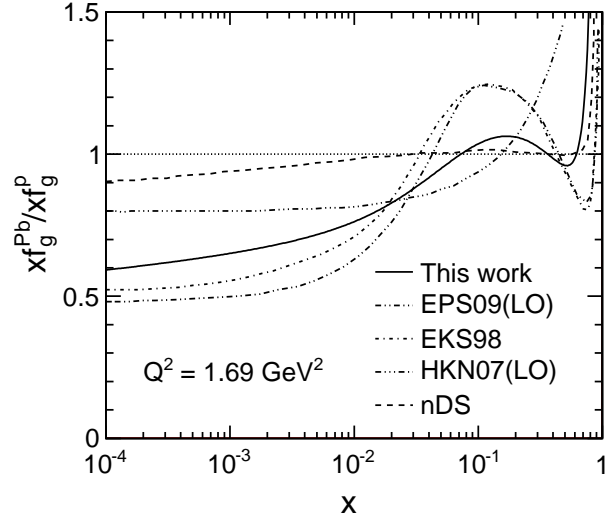}
\caption{Comparisons of our prediction for $xf_g^{Pb}/xf_g^p$ at
$Q^2=1.69~ $GeV$^2$ with the global models: EPS09 \cite{41}, EKS98
\cite{42} (based on the leading-order (LO) global DGLAP analysis),
HKN07 \cite{43} and nDS \cite{44} (next-to-leading-order (NLO) DGLAP
analysis).} \label{fig7}
\end{figure}

   Finally, we present the ratio $F^{Ca}_2/F_2^{2H}$ with various values of $Q^2$ in Fig.8.
One can find that the $Q^2$-dependence of the EMC effect is weak at
$x>0.1$. It implies that it is reasonable to neglect
$Q^2$-dependence in our work.

\begin{figure}[htp]
\centering
\includegraphics[width=0.7\textwidth]{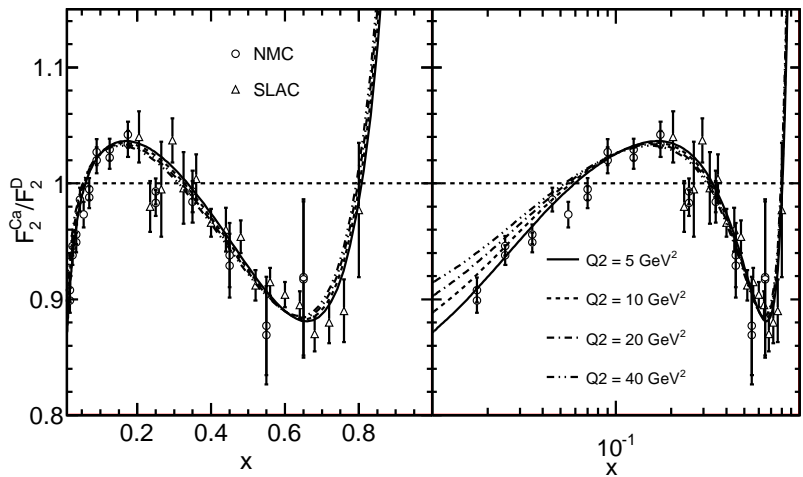}
\caption{$Q^2$-dependence of the EMC effect for $F_2^{Ca}/F_2^D$, (Left) and (Right) take different $x$ scales but for the same results.}
\label{fig8}
\end{figure}

\newpage
\begin{center}
\section{Discussions and Conclusions}
\end{center}

    The main advantage of our explanation of the EMC effect comparing with the other models is that
the parton distributions in both proton and nucleus are
evolved according to a unified dynamics - the DGLAP equation with the ZRS corrections. The
nuclear environment increases nonlinear effects and
deforms the nonperturbative input quark distributions. In
particular, the nuclear shadowing is a natural result of the QCD
evolution rather than the parameterized description as in most of
the EMC effect models. In consequence, the number of free parameters
are much reduced in whole kinematic range. It implies that the
prediction power of our approach for the nuclear parton distributions is enhanced.

       In summary, we obtain new explanation on the EMC effect via the DGLAP equation with the ZRS corrections
and minimum number of free parameters, where the nuclear shadowing
effect is a dynamical evolution result of the equation, while
nucleon swelling and Fermi motion in the nuclear environment deform
the input parton distributions. In consequence,
parton distributions of both proton and nucleus are predicted in
a unified framework. We find that the parton recombination as a
higher twist correction plays an essential role in the evolution of
parton distributions either of proton or nucleus. We show a weak short range
correlation if the Fermi momenta of two nuclei are closed.
In particularly, the nuclear gluon distributions are dynamically
predicted, which are important information for the recherche of the high energy nuclear physics.

\vspace{0.3cm}

\noindent {\bf Acknowledgments}: This work is partly supported by the National Natural Science Foundations of China under the Grants
Number 10875044, 11275120 and Century Program of
Chinese Academy of Sciences Y101020BR0.

\end{document}